\documentclass[CMYK,NoSecthm]{jcomsec_author}  
\usepackage{savesym} 
\savesymbol{AND}     
\usepackage[numbers,sort&compress]{natbib}
\usepackage{multicol}
\usepackage{enumerate}
\usepackage{amssymb}
\usepackage{multirow}
\usepackage{amsmath}
\usepackage{graphicx}
\usepackage{subcaption}
\usepackage{algorithmic}
\usepackage{colortbl}
\usepackage{color}
\usepackage{wrapfig,comment}
\usepackage{array,xcolor,booktabs}
\usepackage{microtype}

\usepackage{xcolor}  

\DisableLigatures[f]{encoding = *, family = * }

\newcolumntype{X}[1]{>{\centering\arraybackslash}m{#1}}

\begin{document}

\begin{frontmatter}



\title{A Q-Learning Approach for Dynamic Resource Management in Three-Tier Vehicular Fog Computing}


\author[DLS]{Bahar Mojtabaei Ranani }
\ead{baharmojtabaei7125@gmail.com \rm(B. Mojtabaei)}
\author[DLS,CorAuth]{Mahmood Ahmadi}
\ead{m.ahmadi@razi.ac.ir \rm(M. Ahmadi)}
\author[DL]{Sajad Ahmadian}
\ead{s.ahmadian@kut.ac.ir \rm{ S. Ahmadian}}
\address[DLS]{Department of Computer Engineering and Information Technology, Razi University, Iran.}
\corauth[CorAuth]{Mahmood Ahmadi.}\
\address[DL]{ Faculty of Information Technology, Kermanshah University of Technology, Kermanshah, Iran}
  
\begin{abstract}
In this paper, a method for predicting the resources required for an intelligent vehicle client using a three-layer vehicular computing architecture is proposed. {This method leverages Q-Learning to optimize resource allocation and enhance overall system performance. } This approach employs reinforcement learning capabilities to provide a dynamic and adaptive strategy for resource management in a fog computing environment. The key findings of this study indicate that Q-learning can effectively predict the appropriate allocation of resources by learning from past experiences and making informed decisions. Through continuous training and updating of the Q-learning agent, the system can adapt to changing conditions and make resource allocation decisions based on real-time information. The experimental results demonstrate the effectiveness of the proposed method in optimizing resource allocation. The Q-learning agent predicts the optimal values for memory, bandwidth, and processor. {These predictions not only minimize resource consumption but also meet the performance requirements of the fog system.} Implementations show that this method improves the average task processing time in compared to other methods evaluated in this study.

\end{abstract}




\end{frontmatter}



\section{Introduction}

Smart transportation and driving have experienced exponential growth in recent years. This is driven by the Internet of Vehicles (IoV), which enhances vehicle connectivity and enables communication with other vehicles and surrounding infrastructure. The IoV has significantly boosted the adoption of other emerging technologies, transforming how we interact with vehicles. For example, autonomous bus fleets in busy environments must interact with other vehicles to prevent accidents and reduce traffic, making real-time decisions without delay. Additionally, in-vehicle entertainment and autonomous driving have revolutionized the driving experience \cite{ref1}.  Undoubtedly, computations such as selecting the optimal route and estimating the distance to the vehicle ahead require resources like the cloud. However, the significant distance between objects and the cloud, along with the high volume of delay-sensitive requests, creates challenges in service delivery. Nevertheless, leveraging the computational capabilities of idle resources near end devices, such as manned or unmanned vehicles, and establishing a fog computing framework known as vehicular fog computing can reduce the number of requests sent to the cloud and decrease response times. Additionally, the offloading process is challenging, and the limited capacity of uplink connections between edge and cloud servers, due to network congestion issues, cannot support the delivery of large-scale content \cite{ref2}.  

 { Nevertheless, resource constraints in vehicular fog computing, along with the dynamic changes and mobility in this system, lead to significant challenges. These include identifying the optimal resource in terms of computational power and ensuring resource availability within the timeframe from request to response. Consequently, intelligent resource allocation in vehicular fog computing requires methods such as machine learning. These methods learn environmental behavior to some extent based on end-device requests and responses, enabling optimal predictions for allocating resources to new requests.}

How can machine learning algorithms be used to optimally predict the resources needed in vehicular fog computing to process a request and provide a response with minimal delay? {To achieve the best prediction of required resources in vehicular fog computing, machine learning algorithms such as hierarchical meta-reinforcement learning can be employed. These algorithms enable processing of requests and delivery of responses with minimal delay.} These algorithms can make optimal resource allocation decisions based on input data and the current state of resources. By training these algorithms with input data and required outputs, improvements in resource prediction and reduced response latency can be achieved. 

 In this paper, a proposed method using Q-learning for managing and predicting the required resources of a vehicles fog computing system with a three-layer architecture was presented. This method offers a dynamic and adaptive approach for resource management in a computational environment, utilizing the capabilities of reinforcement learning.  The three-layer architecture of the vehicular fog computing system is composed of vehicle, fog, and cloud layers, providing a distributed and non-centralized approach to resource management. The vehicle layer uses local resources, while the fog layer acts as an intermediary, offering additional computational and storage capabilities. The cloud layer provides scalable and extensive resources for performing heavier computational tasks.  The results of the experiments demonstrate the effectiveness of the proposed method in optimizing resource allocation. The Q- agent learns optimal actions that minimize resource consumption while meeting the system's performance requirements. This leads to an improvement in efficiency, a reduction in delay, and an increase in the reliability of the system. {The main contributions of this paper are in the following:
 \begin{itemize}
 \item Q-Learning integration in three-tier VFC: Novel application of Q-Learning for real-time prediction of CPU, memory, and bandwidth in a vehicle-fog-cloud architecture, adapting to dynamic vehicular workloads unlike static baselines.
 \item Custom adaptive framework: Introduces a state space encompassing traffic conditions, resource utilization, and app demands, with actions enabling long-term reward maximization through error-learning and multi-objective balancing.
 \item Task probability-based rewards: Pioneers a reward system tied to task arrival probabilities, optimizing allocation to reduce latency and waste, outperforming methods like FCFS and RR in high-mobility scenarios.
 \end{itemize}
  }

The remainder of the paper is organized as follows. Section \ref{relatedwork} presents the related research in vehicular fog computing resource management. Section \ref{proposed} describe the proposed approach for resource prediction in vehicular fog computing systems. 
Section \ref{results} evaluates the proposed solution. Finally, the Section \ref{conclusion} concludes the paper. 

\section{Related Work}
\label{relatedwork}
This section reviews related research in three parts including: The works based on offloading computation to vehicle fog computing nodes, the works based on machine learning in vehicular fog computing, and the works based on predicting the resource requirements in vehicular fog computing.
\subsection{Based on offloading computation to vehicle fog computing nodes}

To incentivize vehicles for task acceptance in fog computing, mechanisms like base station contracts are crucial. These contracts define the relationship between efficiency (computational resources) and reward (payment to vehicles). Vehicles select contracts to maximize profit and act as fog nodes. Task allocation is modeled as a two-sided matching game, considering delay, task size, and cost \cite{ref47}

In \cite{ref48}, machine learning is used for task allocation to nodes in vehicular fog computing. A roadside processing unit (RSU), having learned system performance, advises other units and vehicles. The learning agent interacts with fog nodes to gain performance knowledge. After acquiring complete knowledge, tasks are assigned to nodes with the best performance.

The FOLO method \cite{ref49} is proposed to optimize service latency and reduce quality degradation by allocating tasks to fixed and mobile fog nodes. Formulated as a dual-objective optimization problem (minimizing latency and quality loss), the process includes four stages: discovering mobile fog nodes, sending task requests, allocating tasks to fog nodes, and scheduling task transfers. Fixed fog nodes reside in static infrastructure, while mobile nodes are carried by vehicles with communication modules. The core module manages task allocation and scheduling by tracking node paths and positions.

This approach \cite{ref50} aims to maximize throughput and reduce queue latency using Lyapunov optimization to separate resource allocation and task offloading. An asynchronous federated deep Q-learning algorithm offloads tasks from user vehicles to edge or fog servers. Collaborative vehicular networks, integrating edge and fog computing, enable real-time data processing. Simulations show improved throughput and effective reduction in end-to-end queue latency.

In vehicular fog computing \cite{ref41}, optimizing the connection between roadside processing units and fog servers is key to reducing energy consumption and balancing computational load. The FedDOVe method, using federated deep Q-learning, reduces energy use by up to 49\% and improves load balancing by up to 65\%. It leverages global information without requiring a centralized model. The federated system ensures data privacy while enabling collaborative model development across entities.

Limited computational capacity and power supply hinder long-term vehicle participation in vehicular cloud edge computing networks. In \cite{ref54}, an integrated model optimizes energy consumption and latency in task offloading and resource allocation, formulated as a binary optimization problem. A distributed deep learning approach using parallel neural networks provides near-optimal task offloading decisions. Implementation results demonstrate reduced energy consumption.

\subsection{ Based on machine learning in vehicular fog computing}

In \cite{ref58}, to allocate resources in vehicular fog computing, supervised learning and historical data are used to minimize service latency. The method assigns service requests locally or to a remote cloud server for optimal latency. In dynamic urban environments, a reinforcement learning algorithm with proximal policy optimization detects changes and allocates resources. Roadside units advertise fog services via WAVE messages, allowing new vehicles to join. Fog resource management periodically updates allocations based on demand and status. Evaluations show this approach outperforms conventional resource allocation schemes in service satisfaction.

In \cite{ref59}, deep reinforcement learning with an incentive contract reduces complexity in resource management and allocation. Tasks are rapidly assigned to vehicles with idle resources via deep neural networks. A queue model based on accumulated rewards prevents conflicts from simultaneous task assignments. System performance gradually improves with more vehicles, and mini-batch processing enhances efficiency. Increasing vehicle numbers boosts roadside processing tasks, though idle computational resources also rise. Mini-batch processing reduces computational load and improves model generalization.

This study \cite{ref61} proposes two schemes for joint task assignment and resource allocation in vehicular fog computing. The first scheme divides the problem into two subproblems—task and resource allocation—solved using matching and convex optimization. The second scheme enhances accuracy using 0-1 integer linear programming. Both schemes leverage vehicle mobility predictions to forecast network topology, optimizing task and resource allocation. The goal is to minimize overall latency in vehicular fog computing with a TARA strategy. Simulations show these schemes effectively reduce total latency.

In \cite{ref62}, parked vehicles can serve as fog nodes to process mobile app demands, enhancing wireless network communication. The system comprises three components: a fog controller, parked electric vehicles, and mobile devices. The fog controller manages computational requests, directing them to vehicles or the cloud based on available energy. Requests follow a Poisson process, and vehicle exits follow an exponential distribution. Charge optimization is formulated using a Markov decision process to improve app responsiveness. This approach enhances rewards and battery lifespan through equitable energy allocation.

Vehicular edge computing supports emerging applications in vehicular networks using pervasive edge resources. High vehicle mobility and dynamic network topology complicate task scheduling and resource allocation \cite{ref63}. A matching-based algorithm is proposed to optimize task offloading and resource allocation, reducing latency and costs. It addresses mobility, offloading decisions, resource allocation, and reliable result feedback. Simulations show superior performance compared to existing strategies. The goal is to efficiently utilize resources for low-latency, reliable services.

The paper \cite{ref64} explores vehicle edge computing for automotive services, proposing a framework for intelligent management of heterogeneous resources across vehicles, edge servers, and the cloud. It supports vehicle-to-vehicle and vehicle-to-infrastructure offloading. Resource allocation and task scheduling are jointly optimized to enhance system efficiency and meet user needs. An asynchronous actor-critic algorithm is used for optimal scheduling decisions. The framework aims to maximize resource utilization and improve user service experience. It integrates inter-layer (local, edge, cloud) and intra-layer (among edge servers or users) collaboration

\subsection{Based on predicting the resource requirements in vehicular fog computing}

In \cite{ref67}, by formulating the resource allocation problem, one of the advanced reinforcement learning methods, Actor-Critic, was used to predict the availability of resources in vehicular fog computing. First, the resource allocation problem in vehicular fog computing is formulated to maximize user satisfaction. Then, a set of all divided regions and the collection of vehicular services is presented. Additionally, user satisfaction is defined in terms of service latency and memory space. { The proposed Q-Learning method, a value-based reinforcement learning approach, differs from Actor-Critic algorithms by directly estimating the action-value function (Q-values) to select optimal actions in discrete state-action spaces. This makes it off-policy and tabular, enabling straightforward implementation without requiring separate policy and value estimators. In contrast, Actor-Critic combines an actor (policy network) for action selection with a critic (value network) for evaluation, which supports on-policy learning and handles continuous action spaces more effectively but introduces higher computational complexity and potential instability from policy gradient variance. The key advantage of Q-Learning lies in its simplicity, faster convergence in environments like the three-tier vehicular fog computing architecture described, and lower overhead for real-time adaptation. In such settings, resource allocation (e.g., memory, bandwidth, processor) can be discretized effectively, as evidenced by superior task processing times and service rates in our simulations. In contrast, Actor-Critic methods may excel in scalability for larger, continuous domains, though at the cost of increased training demands. }

The TARA algorithm \cite{ref46} is proposed for predicting resource needs, task allocation to vehicles, and resource assignment in wireless vehicle networks. When a vehicle generates a task difficult to process alone, it sends a request to roadside units for optimized task and resource allocation to minimize delay. The task is divided into N subtasks, each processed by a vehicle with varying file and computational sizes. Multi-hop communication extends the vehicular fog computing service range. The proposed scheme significantly improves performance with larger file sizes compared to others and optimizes resource allocation by reducing bandwidth per link. Simulations confirm that this strategy effectively reduces delay.

The paper \cite{ref68} proposes a deep learning approach for predicting future resource needs. It introduces a resource prediction module and an optimal decision-making module to enhance AI-based resource scheduling. The prediction module is trained with current request and predicted resource data. The decision-making module integrates prediction metrics and HPA to compute optimal container numbers for the API server. HPA in Kubernetes enables automatic pod scaling based on demand. Results show significant reductions in resource occupancy and response latency (over 25\%) in vehicular fog computing.

Resource predicting is essential for fair and efficient allocation. In paper \cite{ref69}, a fog-enhanced vehicle service model is used where vehicles are end-users requesting resources. A probabilistic model predicts fluctuating resource needs accurately. The model categorizes users as privileged and non-privileged. Privileged users receive more resources, with a base price parameter fixed by vehicle type. This approach optimizes resource utilization.

The paper \cite{ref70} models resource prediction in the Internet of Vehicles using a semi-Markov decision process. It introduces a resource reservation strategy and secondary allocation mechanism for adaptive, self-learning dynamic resource allocation. Suitable for high-mobility vehicular environments, it handles service requests at any time. A joint optimal allocation scheme for communication and edge computing resources enhances long-term system revenue. It accounts for variable wireless channel characteristics to improve transmission efficiency and user experience quality. Simulations show that the mechanism reduces service request rejection rates and improves system performance. Table \ref{table1} summarizes the works done in vehicular fog computing resource allocation.

\begin {table*} [!t]

\begin {center}
\caption {Summary of papers on task and resource allocation. }
\label{table1}
\setlength \tabcolsep {6pt}
\begin{small}
\resizebox{0.95\textwidth}{!}{ 
\begin{tabular}{|p{0.08\textwidth}|p{0.44\textwidth}|p{0.28\textwidth}|p{0.28\textwidth}|p{0.1\textwidth}|}
\hline
\rowcolor {blue! 10}

Source & Proposed method & Advantages & Disadvantages & Delay reduction \\ \hline

\cite{ref47}&It proposes an efficient incentive mechanism based on theoretical contract modeling and transforms the task allocation problem into a two-sided matching problem between vehicles and user equipment (UEs).&Reduced network latency, Optimal performance, and Lower complexity& 
With a low number of vehicles, delay increases, and price increases as a solution to competition. & { $ \sim $ 15-20 (inferred from low vehicle count scenarios)} \\ \hline
\cite{ ref49}&This paper presents a novel solution for delay and QoS-optimized task allocation in vehicular edge computing, called the FOLO algorithm. FOLO is designed to support the mobility of vehicles.& It reduces service delay by up to 27\% and improves overall QoS by up to 56\%.& Relatively high computational complexity.
With the selection of complex and random nodes, performance decreases. & {27}\\ \hline

\cite{ref58}& This paper formulates the problem of fog resource allocation using parked vehicles and integrates a proposed algorithm with reinforcement learning to make more efficient resource allocation decisions.&Improving service latency from the users' perspective and minimizing response time for users.  &  Additional delay as well as
the computational load of tasks and
returning results can be prone
to errors from a data delivery perspective. & {$ \sim $ 20 (service latency reduction)} \\ \hline

\cite{ref56}& This paper investigates the problem of energy-aware task allocation in vehicular fog networks using reinforcement learning combined with heuristic information, referred to as URLLC.& It reduces interference between nodes, increases throughput, and ensures the reliability of URLLC V2X communications by selecting the best loading decisions. & Sometimes assumptions can be unrealistic while making specific loading decisions that may not yield the best results. &{N/A} \\ \hline

\cite{ref62} &In this article, the resource allocation problem is formulated as a Markov Decision Process (MDP), and dynamic programming is used to solve the decision-making problem.& Improving communications in a public wireless network, It  performs better than other fixed strategies in terms of reward and energy efficiency.&    Continuous use of the same batteries may significantly reduce the lifespan of electric vehicle batteries. &{N/A}\\ \hline

\cite{ref59} &This paper proposes a contract theory-based incentive mechanism for the current IoV scenario to encourage vehicles to share their idle computing resources. It also uses a combination of contract-based DRL deep reinforcement learning. &Encourage vehicles to share their idle resources. Reduce the complexity of implementing resource management and allocation.
& System performance improves later. &{N/A} \\ \hline
\cite{ref67} &This paper formulates the vehicular fog computing resource allocation problem and uses reinforcement learning to predict the availability of vehicular fog computing resources and service demand. & Maximizes user satisfaction. & Considering the high mobility of vehicles in vehicular fog computing and the rapid change of network topology, the problem of vehicle coordination for vehicular fog computing is a big challenge.  &{N/A}\\ \hline

\cite{ref50}  &  This paper proposes a partial V2V offloading scheme to solve the high mobility of vehicles, which
measures the availability of vehicle services based on the duration of the connection and the diversity of its computational resources.
It also uses a DRL algorithm based on SAC software. & Maximizing the expected reward over a period of time and performing better than other methods.& The highly dynamic vehicle environment that affects the vehicle's task offloading performance does not consider infrastructure as fog nodes. &{N/A} \\ \hline

\cite{ref77} &In this paper, reducing service time by allocating available bandwidth to four types of services is formulated, and network tools are used as a measurement index for allocating available bandwidth. & Accelerate service delivery, save more energy, and improve system survivability. & Power supply limitations of mobile computing devices. &{High ($\sim$ 30-40, accelerates delivery)} \\ \hline

\cite{ref66}  &  This paper presents EnTruVe, an energy-aware virtual machine allocation and TRUst in vehicle fog computing solution that aims to minimize the number of virtual machine migrations and reduce the energy consumption associated with virtual machine processing as much as possible.   & Minimize the number of virtual machine migrations and reduce the energy consumption associated with virtual machine processing. The most energy efficient solution.    &
The number of virtual machine migrations between fog nodes can reduce overall system performance. & {N/A}  \\ \hline

\cite{ref46}   &   This paper, considering the mobility effect in vehicle fog computing, studies the joint task allocation and resource allocation optimization problem, and formulates the joint optimization problem from the Max-Min perspective in order to reduce the overall task delay.   &   Minimizing the delay and also the proposed scheme reduces the bandwidth allocated to each link by keeping the total bandwidth constant and increasing the number of vehicles.   &

This scheme does not provide significant performance improvements when the file size is small. & {High ($\sim $ 40, min delay)} \\ \hline

\cite{ ref63}   &   The problem of shared task loading and resource allocation in vehicular networks is studied to optimize the system tools in order to reduce the latency and cost of computing and communication services. To solve this problem, a matching-based task loading and resource allocation algorithm is proposed.   & It minimizes the latency and cost of using heterogeneous communications and computing resources located in roadside processing units and vehicles.   &   Does not consider security and privacy. & {High ($\sim$ 35, min latency/cost)} \\  \hline

\cite{ref64} & Joint workload computing and demand-based resource allocation in automotive edge computing for automotive services and applications are investigated.  & Maximizes resource utilization.
Ensures a diverse user experience in vehicle edge computing.    & 
The number of vehicles is limited and resource allocation may be delayed as the number of vehicles increases. &{N/A} \\ \hline

\cite{ref65}    &  To review and present a new algorithm for allocation and optimization of exploration resources in a MultiFog-Cloud environment.   & Optimize runtime and communication latency.
Minimize cloud load.  & Does not optimize revenue. Does not consider mobile users’ preferences. &{ Medium ($\sim$ 20, runtime/latency)} \\ \hline
{Proposed approach} &{ Q-Learning-based reinforcement learning in a three-tier (vehicle-fog-cloud) architecture for real-time prediction of CPU, memory, and bandwidth.}&{Adaptive to dynamic vehicular mobility and workloads, learning optimal allocations from experience without predefined rules.} & {Q-learning growth in large state-action spaces may increase memory overhead on resource-constrained fog nodes. Sensitive to reward weighting; imbalances could prioritize one objective}  & { $ \sim $ 35-45 (inferred from APT graphs)} \\  \hline

\end {tabular}
}
\end{small}
\end {center}
\end {table*}

\section{Proposed Method}
\label{proposed}

 { Achieving an efficient resource prediction solution is a challenging task in resource management. Ideally, this solution should, first, avoid resource wastage and, second, ensure that resources are fully sufficient for task allocation. Such requirements are particularly demanding in environments with time-varying workloads.} The use of reinforcement learning in the proposed method for predicting required resources in computing (such as CPU allocation, memory, and bandwidth) is driven by several key reasons.
\begin{itemize}
\item The dynamic and uncertain nature of systems: Computational environments are typically dynamic and uncertain. Resource demands can continuously change, and traditional algorithms may not optimally cope with these variations. Reinforcement learning enables an agent to learn and gradually improve resource management policies through direct interaction with the environment. This is particularly significant in cloud systems and large-scale infrastructures.

\item  Ability to adapt to environmental changes: Reinforcement learning enables agents to improve optimal resource allocation policies over time by accumulating experiences. As conditions change, such as varying processing loads or increasing/decreasing resource demands, the agent can dynamically adapt.
\item Long-term optimization: In reinforcement learning, the goal is to enable the agent to maximize long-term rewards rather than making short-term decisions. In resource management, a decision that appears optimal in the short term may lead to issues in the long term (e.g., resource overloading). Reinforcement learning can strike a balance and make decisions that are more optimal in the long run.

\item  Addressing complexity and high scalability: Managing resources in large and complex systems requires algorithms capable of optimizing in the face of vast amounts of data and diverse states. Reinforcement learning can handle high complexity and large state spaces (e.g., various combinations of workloads and resource demands) more effectively than traditional methods.
\item Learning from errors: Reinforcement learning allows agents to learn from their errors and adjust their policies accordingly. In resource allocation problems, this can lead to continuous improvement of system performance and prevent the repetition of previous errors.
\item Solving multi-objective problems: In many cases, optimal resource allocation must simultaneously consider multiple criteria, such as reducing latency, minimizing energy consumption, or lowering economic costs. Reinforcement learning can optimize these multi-objective problems concurrently.
As a result, due to its adaptability, long-term optimization capabilities, and dynamic learning, reinforcement learning is a suitable choice for the problem of prediction and resource allocation in computational environments.
\end {itemize}

\begin {figure} [!t]
\begin{center}
\subfloat [A general architecture of VFC.] {\includegraphics [scale = 0.3] {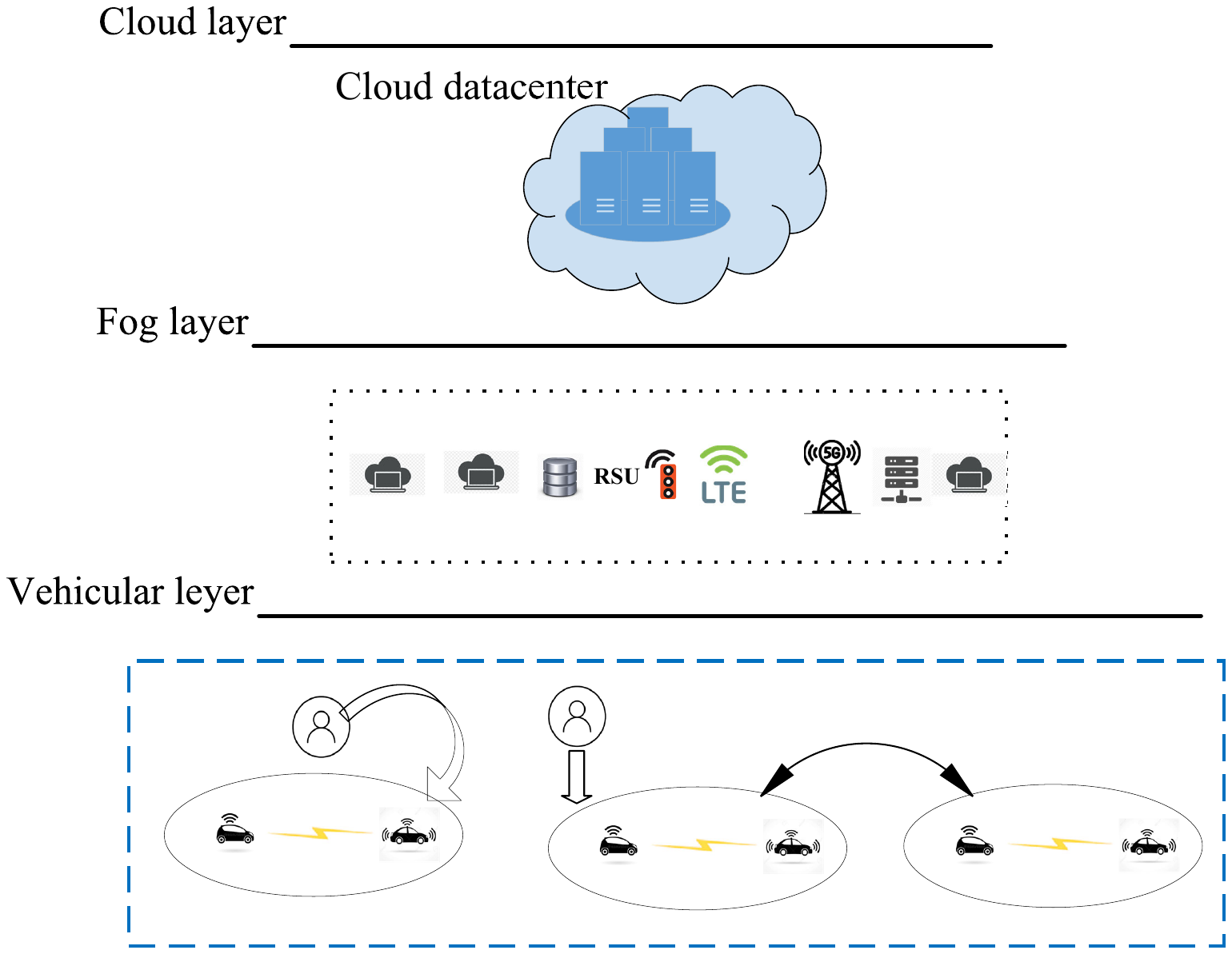} \label {sensetie021} }
\qquad
\subfloat [An implementation of VFC.] {\includegraphics [scale=0.4] {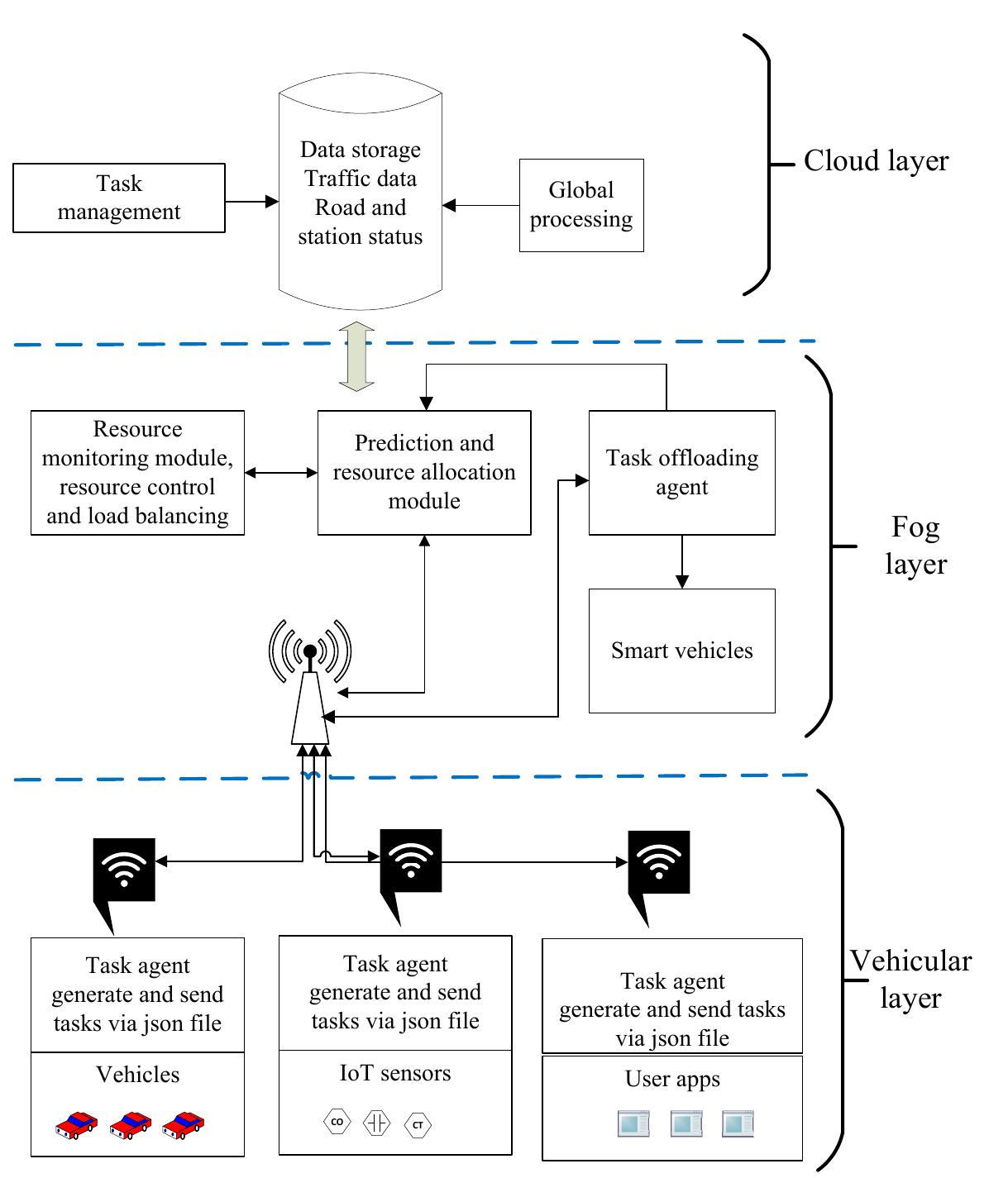} \label {sensetie022}}
\caption {A general architecture VFC and its implementation (a) A general VFC architecture. (b)An VFC implementation. }
\label {fig1}
\end{center}
\end {figure}

Figure \ref{fig1}(a) shows the implementation of a vehicle fog computing architecture. The figure depicts the various components of the system and how they interact.
According to Figure \ref{fig1}(a), the vehicular fog computing architecture consists of the following layers: vehicle layer, fog layer, and cloud layer. The vehicle layer represents the vehicles in the network, the fog layer includes local fog nodes, and the cloud layer represents centralized cloud data centers. The design and implementation of the network is based on appropriate settings and connections between layers. To design and implement the vehicular fog computing architecture, each layer is considered separately, and the settings and connections of the layers are defined. It should be noted that the arrows between the layers indicate the assignment of the task to a higher layer for processing.

Vehicle Layer: The vehicle layer represents the vehicles in the network. Each vehicle will be equipped with communication and computing capabilities. The configuration of the vehicle layer tasks includes the following:
\begin{itemize}

\item Communication technology: Vehicles may use technologies such as Wi-Fi or cellular networks to communicate.
\item Computing capabilities: Each vehicle will have built-in computing resources to process and analyze data.
\item Vehicle-to-vehicle communication (vehicular communications): Vehicles must be able to communicate with each other to share information and coordinate their actions.
\item Vehicle-to-infrastructure communication: Vehicles must communicate with roadside infrastructure such as traffic lights or road sensors to collect data in real time.
\end{itemize}

Fog layer: The fog layer consists of local fog nodes located in the vicinity of vehicles. These nodes provide additional computing and storage resources to support latency-sensitive and real-time applications. The fog layer configuration includes the following:

\begin{itemize}

\item Fog node location: Fog nodes should be strategically placed to ensure coverage and minimize latency. They can be placed in areas with high vehicle density or critical infrastructure points.
\item Connectivity: Fog nodes should be connected to the vehicle layer via reliable, low-latency communication links such as wired connections or high-speed wireless networks.

\item Computing and storage resources: Fog nodes must have sufficient processing power and storage capacity to handle the computing needs and deployed applications.
\item Load balancing: Multiple fog nodes can be deployed to distribute the computing load and provide fault tolerance.
\end{itemize}

Cloud layer: The cloud layer represents centralized cloud data centers that provide global-scale computing and storage capabilities. The cloud layer configuration includes the following:

\begin{itemize}
\item Cloud data center location: Cloud data centers should be strategically located to provide efficient access to fog nodes and minimize network latency. 
\item Connectivity: Cloud data centers should be connected to the fog layer using high-capacity network links, such as fiber optic connections, to handle the large volume of data generated by vehicles.
\item Compute and storage resources: Cloud data centers must have powerful compute infrastructure and large-scale storage facilities to support intensive applications and perform complex data analytics. 
\item Scalability and elasticity: Cloud data centers must be scalable to handle increasing demand and provide elasticity to dynamically allocate or release compute resources based on workload.
\end{itemize}

Figure \ref{fig1}(b) shows the implementation aspects of a resource allocation and predicting system using a reinforcement learning algorithm, which includes various components for task management and data processing. The following describes each component and their interactions:
\begin{itemize}
\item Vehicular layer: This layer includes users, objects, and applications that generate tasks and send each task to the roadside units in the form of a JSON file. The content of the JSON file is data fields such as resources and their number, executable code, parameters sent, parameters received, etc. Below are a few examples of entities in the customer layer briefly described:
\begin{enumerate}

\item Vehicles: These vehicles generate tasks according to their needs, smart cars may interact with IoT sensors to collect environmental data and make better decisions based on it.

\item IoT Sensors: IoT sensors collect data related to the environment, vehicle status, etc. and send the relevant tasks to the system. This data can include parameters such as speed, geographical location, temperature, humidity, etc.
After collection, this data is sent to the computing layer at the edge. This communication is usually done through wireless networks. In the computing edge layer, the data received from the sensors is processed. This processing includes initial analysis, data filtering, and immediate decision-making that does not require sending the data to the cloud. After initial processing, the information can be used to assign tasks to different nodes.
\item User applications: Applications act as the user interface in distributed computing systems and play a key role in data collection and task generation. These programs collect information from users and their surroundings, such as location and specific needs, and based on that, create tasks for processing. {After sending these tasks to the edge and cloud computing layers, the processed results are returned to the applications for display to users. This improves the user experience and provides more accurate information.}
\end{enumerate}

\item Fog layer: This layer deals with local processing, task management, node resource management, and task resource prediction and includes thefollowing modules:

\begin{enumerate}
\item Task offloading agent: This module receives tasks from the client layer and manages them. First, the agent in this module collects and analyzes information about the resource status of the nodes, their current workload, and their capacity. Then, based on this analysis, the resource predicting and allocation module estimates the amount of resources required and decides which node is the most suitable option for processing the task. {After selecting a node, the agent is responsible for transferring the tasks to the destination nodes and monitoring their execution performance. It also calls the resource monitoring module to update the resource table for those nodes.} This process helps optimize system performance and ensure efficient resource distribution. Finally, it returns a feedback of the waiting time and execution time of the tasks to the predicting and allocation module for optimization.
\item  Smart cars with processing resources: These cars act as server computing nodes in the environment and are responsible for executing tasks or sending them to other resources for analysis. Smart cars perform a variety of tasks and act as executive units in the system. These cars are able to collect data and provide feedback to the system. After tasks are assigned in the computing layer, information about how to perform these tasks and the resources required are sent to the smart cars so that they can effectively perform their tasks.
\item Resource prediction and allocation module: {This module employs the Q-Learning algorithm and data analysis techniques to examine feedback from executed tasks, workload patterns, and resource consumption trends. The aim is to generate accurate predictions of future task requirements.} After analyzing the data, this module allocates resources optimally to tasks and decides how much resources to consider for executing tasks to maximize system efficiency. This process gradually reduces latency, increases efficiency and improves workload between different nodes in the system.
\item Resource monitoring module: This module balances the workload of VFC layer nodes by monitoring the amount of resources used and free for each node. Resource monitoring is a key tool for effective management of computing systems. This process involves tracking and recording the real-time usage of each node's processor, memory, bandwidth, and storage to prevent problems such as overloading and insufficient resources. By analyzing the collected data, performance optimization and resource reallocation can be done, and also using historical data to predict future needs.
\end{enumerate}
\item Cloud layer: This layer handles big data processing and storage of related information:
\begin{enumerate}
\item Task management: A central location for storing information about traffic, road conditions, and stations, and for receiving management and scheduling.
\item Global processing: Analyzing data on a large scale and performing complex calculations.
\end{enumerate}
\end{itemize}
\subsection{Problem modeling}
This section defines the variables and factors that are useful in the resource allocation decision-making process. These factors include the utilization of available resources, traffic conditions, and application needs. The model features a state space comprising a set of states and a set of actions for each state. When an agent performs an action, it transitions from one state to another. Executing an action in a particular state yields a reward (a numerical score) for the agent.

\subsubsection{State space}
A state space is a multidimensional space in which each dimension represents one of the variables. The values within each dimension can be discrete or continuous, depending on the nature of the variable. Usually, one variable is not sufficient for each of the factors of traffic conditions, current resource usage, and resource availability. Usually, multiple variables are needed to accurately represent and capture the complexity of these factors. For example, traffic conditions can include variables such as the number of concurrent requests, the type of applications, and the expected demand for each type. Similarly, current resource utilization may consist of multiple variables such as CPU utilization, memory utilization, and network bandwidth utilization. Resource availability can also include variables such as the number of available compute nodes, the amount of free disk space, and network capacity. By considering multiple variables for each factor, a state space can provide a more comprehensive representation of the system state for decision making. Some of these important variables and factors that may play a role in resource allocation decisions include:
\begin{itemize}
\item Current resource utilization: This refers to the current usage levels of various resources such as CPU, memory, disk space, or network bandwidth. It provides information about the amount of resources currently being consumed by the system.
\item Traffic conditions: This variable considers the current workload of the system such as the number of requests, application type, and expected demand. This helps in adjusting the resource allocation based on the current system load.
\item Application requirements: These are the specific needs or demands of the application running on the system. This includes factors such as response time, service level agreements, throughput requirements, or any other application-specific criteria.
\item Resource availability: This factor reflects the availability of resources in the system, taking into account capacity and any potential constraints. This includes information about the number of available compute nodes, storage devices, or any other resources in the infrastructure.

\end{itemize}
Table \ref{table2} depicts the state space variables and their values.

\begin{small}
\begin{table}[!h]
\centering
\caption{System variables and their values}
\label{table2}
\resizebox{0.5\textwidth}{!}{ 
\begin{tabular}{lll}
\toprule
\multicolumn{3}{c}{\textbf{Current resource usage}} \\
\midrule
Variable & Meaning & Value \\
\midrule
CU & CPU usage & \{Low, medium, high\} \\
MU & Memory usage &\{Low, medium, high\} \\
DSU & Disk space usage & \{Low, medium, high\} \\
NBU & Network bandwidth usage & \{Low, medium, high\}  \\
\bottomrule
\\
\multicolumn{3}{c}{\textbf{Traffic conditions}} \\
\midrule
Variable & Meaning & Value \\
\midrule
NR & Number of requests & \{Low, medium, high\} \\
AT & Application type & \{Light, medium, heavy \} \\
ED & Expected demand & \{Low, medium, High\} \\
\bottomrule
\\
\multicolumn{3}{c}{\textbf{Application requirements}} \\
\midrule
Variable & Meaning & Value \\
\midrule
RT & Response time & \{Fast, medium, slow\} \\
SLA & Service level agreement & \{Fulfilled, not fulfilled\} \\
OR & Operational requirements &\{Low, medium, high\}\\
\bottomrule
\\
\multicolumn{3}{c}{\textbf{Resource availability}} \\
\midrule
Variable & Meaning & Value \\
\midrule

NCN & Number of computation node &\{Low, medium, high\} \\
ASD & Availability of storage devices & \{Low, medium, high\} \\
\bottomrule
\end{tabular}
}
\end{table}
\end{small} 

The total number of variables will be the sum of the variables in each category, which includes current resource usage, traffic conditions, application requirements, and resource availability. In this case, there are 12 variables. For example, if CPU usage, memory usage, and disk space usage are considered as relevant variables, a three-dimensional state space is defined.

\begin{equation}
\begin{aligned}
S = \{ (X_0, X_1, X_2) \ \mid\ 
&0 \leq X_0 \leq M_1,\\
&0 \leq X_1 \leq M_2,\\
&0 \leq X_2 \leq M_3 \}
\end{aligned}
\label{Eq1}
\end{equation}

In Eq. \ref{Eq1}, $S$ represents the state space, $X0$, $X1$, and $X2$ represent the CPU utilization, memory utilization, and disk space utilization, respectively. $M1$, $M2$, $M3$ are the maximum utilization values for each resource. This state space can be used to evaluate different resource allocation decisions and find the optimal allocation based on the current state of the system. By analyzing different points in the state space, the best resource utilization can be determined by considering other factors.
\subsubsection{Action space}
This section specifies the actions that can be performed in the resource allocation process. These actions can include increasing or decreasing resources, transferring computational tasks, or offloading tasks between layers.  Action space refers to the set of all available operations that can be performed in resource allocation. In the context of resource allocation in computing systems, action space includes various operations such as increasing or decreasing resources, transferring computational tasks, and offloading tasks between different layers or components.

\subsubsection{ Reward function}
A reward function is designed that reflects the goals and objectives of the system. This could include minimizing resource waste, maximizing resource utilization, reducing response time, or maintaining a desired quality of service. The reward function should guide the reinforcement learning process toward desired behaviors and outcomes. The reward function in reinforcement learning minimizes the immediate desirability of a particular state or action taken by an agent in an environment. This is crucial in guiding the agent's learning process and incentivizing it to achieve the system's goals and objectives. The specific design of the reward function depends on the nature of the problem and the desired behaviors and outcomes. 

{ The reward function are specified based on the defined goals. The goals are minimizing resource wastage, maximizing resource utilization, reducing response time and maintaining quality of service. Therefore, the reward function can be defined based on Eq. \ref{Eq21}. The value of the weights $w_1$, $w_2$, $w_3$ and $w_4$ are determined in trial and error manner and are set to the 0.3, 0.3, 0.2 and 0.2.}

\begin{equation}
{
\begin{aligned}
r(s,a) &= -w_1 \cdot \text{resource}_{\text{wastage}}(s,a) \\
       &\quad + w_2 \cdot \text{resource}_{\text{utilization}}(s,a) \\
       &\quad + w_3 \cdot \text{response-time}(s,a) \\
       &\quad + w_4 \cdot \text{qos}(s,a)
\end{aligned}
}
\label{Eq21}
\end{equation}
 {In Eq.\ref{Eq21}, the first term of reward is calculated as the negative product of the weighting coefficient $w1$ and the amount of resource waste. By penalizing wasteful actions, the agent is encouraged to minimize resource use. The weighting factor $w_1$ is determined when implementing a reinforcement learning system. This indicates the importance or priority given to minimizing resource waste. The implementation is chosen based on their understanding of the system and their desired trade-offs between resource use and wasting some amount of $w_1$. A higher value of $w_1$ places more emphasis on minimizing waste, while a lower value gives more priority to other aspects of the system's goals. The specific value of $w_1$ can be adjusted through trial and error to find the optimal balance for the system. 
 The coefficient $w_2$ is the scaling factor $k$ and is determined through trial and error during implementation.}
  Here, the goals that are achieved by the reward function are:

\begin{enumerate}
\item Minimizing resource waste:

\begin{equation}
{
\begin{aligned}
\text{resource}_{\text{wastage}}(s,a) &=  \Bigg[ \sum_{i=0}^{n} \left( A_C - E_C \right) \\
&\quad + \sum_{i=0}^{n} \left( A_M - E_M \right) \\
&\quad + \sum_{i=0}^{n} \left( A_B - E_B \right) \Bigg]
\end{aligned}}
\label{Eq3}
\end{equation}

\begin{itemize}
    \item \( A_C \): Actual CPU usage
    \item \( E_C \): Efficient CPU usage
    \item \( A_M \): Actual memory usage
    \item \( E_M \): Efficient memory usage
    \item \( A_B \): Actual bandwidth usage
    \item \( E_B \): Efficient bandwidth usage
\end{itemize}


Actual CPU utilization refers to the current level of CPU utilization in a system, indicating how much processing power is being used at any given time. It represents the real-time demand for CPU resources by running processes or programs. To calculate resource wastage, it can be determined by measuring the difference between the actual resources used and the optimal or expected resources required for a given task or operation. This can involve comparing factors such as CPU utilization, memory utilization, network bandwidth, and power consumption against thresholds. For example, if a fog system is responsible for processing a set of tasks, resource wastage can be calculated by summing the differences between the actual CPU utilization for each task and the CPU utilization that was sufficient to efficiently complete the tasks. The same approach can be applied to other resources such as memory or network bandwidth.

\item Maximize resource utilization: When considering different types of quantities such as CPU usage, memory usage, and network bandwidth usage, it may not make sense to simply summarize them in a resource usage equation. This is because these quantities may have different units of measurement and scales, which can lead to inaccuracies in the overall calculation. Instead, a more appropriate approach is to normalize each of these quantities individually before multiplying them by the weight. Normalization involves scaling the values of each quantity to a standard range, usually between 0 and 1, so that they can be directly compared and combined in a meaningful way. The equation for normalizing resources with weighting factors can be expressed in Eq. \ref{Eq5}.

\begin{equation}
{
\begin{aligned}
 resource_{Utilization}(s,a) = \Bigg(&
   w_{2,1} \cdot \sum_{i=0}^{n}  NCU + \\
    & w_{2,2} \cdot \sum_{i=0}^{n} NMU + \\
    & w_{2,3} \cdot \sum_{i=0}^{n} NNBU
    \Bigg)
\end{aligned}}
\label{Eq5}
\end{equation}

\begin{itemize}
\item NCU=normalize(CPU usage)
\item NMU=normalize(memory usage)
\item NNBU=normalzie(network bandwidth usage)
\end{itemize}

{Where, $w_{2,1}$, $w_{2,2}$, $w_{2,3}$ are the weights assigned to each normalized quantity,  and are set to 0.4, 0.3, and 0.3, respectively}. The $normalize()$ is a function that scales the values of each quantity to a standard range (e.g. between 0 and 1).


The goals of minimizing resource waste and maximizing resource utilization are not the same. Minimizing resource waste focuses on reducing the amount of resources that are underused or not used. The goal is to optimize resource allocation and prevent unnecessary waste, which can help conserve resources and reduce costs.

\item Reduce response time: Here, the reward {response-time} is inversely proportional to the response time. $T_{current}$ represents the current response time and $T_{Max}$ is the maximum desired response time. The closer the response time is to the desired limit, the higher the reward, encouraging the agent to reduce the response time.
\begin{equation}
{ response-time(s,a)=\frac{T_{max}-T_{current}}{T_{max}}}
\label{Eq7}
\end{equation}

According to Eq. \ref{Eq7}, as the response time $T_{current}$ approaches the maximum desired response time $T_{max}$, the amount of reward increases. This encourages the agent to reduce the response time even further in order to receive more reward; therefore, the higher the reward, the closer the response time is to the desired limit. 

\item Maintaining the desired quality of services: The {qos} reward is calculated based on the difference between the achieved quality and the desired quality of the service. {The coefficient $\frac{1}{e^{*}}$} is a negative exponential function used to assign higher rewards for achieving quality close to the desired level. This guides the agent to maintain the desired quality throughout his decision-making process.

\begin{equation}
{qos(s,a)=\frac{1}{{e}^{\left( quality_{desired}-quality \right)}}}
\label{Eq8}
\end{equation}

{The function $e^{-(*)}$,} is actually a reduction factor that is used to reduce the maximum reward value. This reduction can be used to adjust the reward value to a desired value or to balance the problem. Quality is measured based on factors such as latency, throughput, and reliability.

In this case, quality can be calculated as the weighted average of these factors using appropriate weights:

\begin{equation}
 {
\begin{aligned}
quality &= w_{3,1} \cdot \frac{1}{latency} + w_{3,2} \cdot throughput \\
&\quad + w_{3,3} \cdot reliability
\end{aligned}}
\label{Eq9}
\end{equation}

{$w_{3,1}$, $w_{3,2}$, and $w_{3,3}$ are weights assigned to each factor that reflect their importance in determining overall quality.} These weights can be determined based on specific system priorities.

It is a measurement of the number of units of information that can be processed in a given time, usually measured in bits per second. It indicates the rate at which data can be successfully transmitted or processed in a system. The parameter reliability refers to the percentage of requests that are successfully processed without encountering failure errors. This indicates the reliability of the system in achieving the expected results.

\end{enumerate}

The four objective functions do not add up directly. Instead, they represent different aspects and goals of the system. Each objective function focuses on a specific goal and encourages the agent to optimize its decisions for the base.

The first objective function, 'minimize resource waste', penalizes wasteful actions and encourages the agent to minimize resource usage. Its goal is to optimize resource allocation and prevent unnecessary waste, ultimately conserving resources and reducing costs.

The second objective function, 'maximize resource utilization', encourages the agent to fully utilize available resources to maximize efficiency. It prioritizes efficient resource utilization, thus maximizing system performance or throughput.

The third objective function, 'reduce response time' rewards the agent for reducing response time. This function encourages the agent to prioritize actions that lead to faster processing and improved system responsiveness.

The four objective function 'maintaining the desired quality of service' aims to incentivize the agent to maintain the desired level of quality in terms of latency, throughput, and reliability. It allocates higher rewards for achieving quality close to the desired level.

The impact of these objective functions on the system is not the same, because each function emphasizes a different aspect and goal. For example, minimizing resource waste and maximizing resource utilization may have contradictory effects. Minimizing resource waste may lead to underutilization of resources, while maximizing resource utilization may lead to overutilization of resources; therefore, finding the optimal balance between these goals is very important.

\subsubsection{QL reinforcement learning agent training}
The proposed model is run on all nodes of the fog layer. The collected training data, defined state space, action space, and reward function are used to train the QL reinforcement learning agent. One of the most popular reinforcement learning algorithms is the QL reinforcement learning algorithm, which uses experience to learn to update the value of actions in a given situation, known as the Q-value.

The QL reinforcement learning algorithm is specifically used for Markov reinforcement control problems where the agent operates in a well-defined environment and its goal is to learn an optimal strategy that maximizes the reward. In this algorithm, the Q-value for each state-action pair is updated as feedback from past experiences. This update is usually done using a QL reinforcement learning update equation.

The algorithm and pseudocode of the training process are given in Algorithm \ref{algo1}.

\begin{algorithm}
\caption{Q-Learning Algorithm}
\begin{algorithmic}[1]
\STATE Import numpy as np
\STATE \textbf{function} initialize\_q\_values(num\_states, num\_actions)
\STATE \quad return np.zeros((num\_states, num\_actions))
\STATE \textbf{end function}
\STATE \textbf{function} epsilon\_greedy\_action\_selection(Q\_values, state, epsilon)
\STATE \quad \% Implementation of epsilon-greedy action selection strategy
\STATE \quad \% Return selected action
\STATE \quad \textbf{pass}
\STATE \textbf{end function}
\STATE \textbf{function} execute\_action(action)
\STATE \quad \% Execute the selected action in the environment
\STATE \quad \% Return next state and reward
\STATE \quad \textbf{pass}
\STATE \textbf{end function}
\STATE \textbf{function} update\_q\_value(Q\_values, state, action, next\_state, reward, learning\_rate, discount\_factor)
\STATE \quad \% Update Q-value for the current state-action pair using the Q-learning update rule
\STATE \quad \textbf{pass}
\STATE \textbf{end function}
\STATE Initialize Q-values
\STATE Q\_values = initialize\_q\_values(num\_states, num\_actions)
\STATE \% Loop over episodes
\FOR{episode in range(num\_episodes)}
    \STATE \% Display current system status
    \STATE display\_system\_status()
    \STATE \% Loop over time steps
    \FOR{time\_step in range(max\_time\_steps)}
        \STATE \% Decide on an action using an exploration-exploitation strategy
        \STATE action = epsilon\_greedy\_action\_selection(Q\_values, current\_state, epsilon)
        \STATE \% Execute selected action in the environment
        \STATE next\_state, reward = execute\_action(action)
        \STATE \% Update Q-value for the current state-action pair using Q-learning update rule
        \STATE update\_q\_value(Q\_values, current\_state, action, next\_state, reward, learning\_rate, discount\_factor)
        \STATE \% Move to the next state
        \STATE current\_state = next\_state
    \ENDFOR
\ENDFOR
\end{algorithmic}
\label{algo1}
\end{algorithm}

This pseudocode contains separate functions for each operation, including initializing Q-values, selecting the action using the epsilon-greed strategy, executing the action in the environment, and updating the Q-value for the current state-action pair. First, the Q-Values for each pair of states and actions are randomly initialized with a predetermined initial value. Then, in the second step, for each part, the following steps are performed: Displaying the current state of the system, looping over time steps, selecting an action using an exploration and exploitation strategy, executing the selected action in the environment, observing the next state and reward, updating the Q value for the current state and action pair using the Q-learning update rule, and moving to the next state.

In the third step, the functions used in this pseudocode, such as $initialize\_q\_values()$ and $update\_q\_value()$, must be implemented separately to perform the corresponding operations.

The training algorithm using Q-learning is as follows: Input: Number of training iterations (e.g., 1000 iterations), learning coefficient (e.g., 0 and 1), exploration parameter (e.g., 0 and 3), number of states and actions, and Q-value update equation. Output: Q-value table for each state and action pair.

Lines 1 to 3: Define the function $initialize\_q\_values$ which returns a zero-filled array of dimensions (number of states, number of actions) that holds the Q-values. Line 5 defines the {$epsilon\_greedy\_action \_selection$} function which implements the epsilon-greedy action selection strategy.  The function returns an action that inputs the $Q\_values$ entity, the current state, and the epsilon value. Lines 8 to 10 define the $execute\_action$ function that executes the selected actions in the environment and returns the next state and reward. Lines 12 to 14 define the $update\_q\_value$ function that updates the Q-value for the current state-action pair using the Q-learning update rule. Line 16 initializes the Q-values using the $initialize\_q\_values$ function. Lines 18 to 27 are the main loop of the algorithm, which consists of iterating over episodes and time steps. Line 21 selects the action using the epsilon-greedy strategy, line 23 performs the selected action in the environment and obtains the next state and reward, line 25 updates the Q-value based on the Q-learning update formula, line 27 shows the transition to the next state as the current state.

These steps are generally considered a reinforcement learning process in which an agent learns how to act in a specific environment. This process includes initialization, exploration and exploitation, action execution, updating the Q value, and transitioning to the next state. These steps continue as an iterative cycle until the agent converges to the desired learning or a certain number of training iterations are performed. During the training process of a QL reinforcement learning agent, the agent interacts with the environment by performing actions based on the current state and receives feedback in the form of rewards or penalties. The agent aims to learn the optimal strategy to maximize the cumulative reward over time. At each time step, the agent observes the current state of the system, selects an action based on its policy (which can be heuristic or greedy), and executes the action. After performing the action, the agent receives a reward based on the outcome obtained from the environment. The reward is then used to update the Q-value associated with the state-action pair using the Q-learning update rule equation:

\begin{equation}
Q(s, a) = Q(s, a) + \alpha \left[ R + \gamma \max(Q(s', a')) - Q(s, a) \right]
\label{Eq10}
\end{equation}

$ Q(s, a) $: The current estimated value of taking action $ a $ in state $ s $.
$ \alpha $: The learning rate (a value between 0 and 1), which determines the step size of the update.
$ R $: The immediate reward received after taking action $ a $ in state $ s $.
$ \gamma $: The discount factor (between 0 and 1), which weights the importance of future rewards. $ \max(Q(s', a')) $: The maximum estimated value of the next state $ s' $ over all possible actions $ a' $.
$ [R + \gamma \cdot \max(Q(s', a')) - Q(s, a)] $: The temporal difference (TD) error, which measures the difference between the current estimate and the new estimate based on the reward and the best future action.

This update rule adjusts the $ Q $-value based on the difference between the predicted reward and the actual reward plus the discounted value of the next best action, allowing the agent to learn an optimal policy over time.

After the QL reinforcement learning agent is trained on all nodes in the fog layer, it is ready to make decisions and optimize the defined objective functions in real time. The agent uses its learned knowledge and experiences to assess the current state of the system, consider available resources, and make decisions about task scheduling and resource allocation. { The differential learning algorithm optimizes the objective functions by continuously evaluating the current state, taking actions based on its learned policy, receiving feedback in the form of rewards or penalties, and updating its knowledge for errors.} The Q-learning algorithm updates the Q values based on the rewards received and uses these updated values to make better decisions in the future.  The training and optimization process is performed on each individual node of the fog layer. Each node acts independently to make decisions based on its local observations and interactions with the environment. There is no need for a central server to coordinate the decision-making process, as each node is able to learn and optimize its actions independently. 
{In general, the Q-Learning agent optimizes the defined objective functions by continuously learning and adapting to changing task requirements and system conditions. Its ultimate goal is to maximize resource utilization, minimize waste, reduce response time, and maintain the desired quality of service in the vehicular fog system.}

\section{Evaluation Results}
\label{results}
This section presents the evaluation results of the proposed solution.
\subsection{ Dataset}

Real vehicle routes are used as traffic input for the simulation. These routes are obtained from the Didi GAIA \footnote{ Https://github.com/Whale2021/Dataset} open dataset, specifically from a $3\times3$ $km^2$ area in Qingyang District, Chengdu, China, on November 16, 2016 \cite{ref81}. According to the dataset, four different service scenarios with different time periods are considered.

Various statistics related to the traffic data are summarized in { Table \ref{table3}.} These statistics include the total number of vehicle trackings, the average dwell time (ADT) in the $3 \times 3$ $km^2$ area, the variance of the dwell time (VDT), the average number of vehicles (ANV) per second, the variance of the number of vehicles (VNV), the average speed of vehicles (ASV), and the variance of the speed of vehicles (VSV).

\begin{table}[h!]
\centering
\caption{Traffic data characteristics.}
\resizebox{0.5\textwidth}{!}{ 
\begin{tabular}{|c|c|c|c|c|c|c|c|c|}
\hline
\multicolumn{9}{|c|}{Traffic data} \\
\hline
Scenario & VSV & ASV & VNV & ANV & VDT & ADT & Trace & Time \\
\hline
NO.1 & 2.61 & 5.22 (m/s) & 11.6 & 474.6 & 123.8 & 198.3 (s) & 718 & 8:00 - 8:05 (h) \\
\hline
NO.2 & 2.73 & 5.59 (m/s) & 5.38 & 541.6 & 125.1 & 188.5 (s) & 862 & 13:00 - 13:05 (h) \\
\hline
NO.3 & 2.40 & 4.60 (m/s) & 7.76 & 608.0 & 122.5 & 196.5 (s) & 928 & 18:00 - 18:05 (h) \\
\hline
NO.4 & 3.16 & 7.30 (m/s) & 3.93 & 207.9 & 124.1 & 173.7 (s) & 359 & 23:00 - 23:05 (h) \\
\hline
\end{tabular}
}
\label{table3}
\end{table}

Certainly for short-run tasks, the time is 193. This is because the analysis was done over short intervals and specific time periods. In this case, 5-minute intervals were chosen to capture specific traffic conditions or patterns at those times.

Scenario 1, which occurred from 8:00 to 8:05, involved a total of 718 vehicles tracked in a $3 \times 3$ $km^2$ area in Qingyang District. The average dwell time of these vehicles was 198.3 seconds with a variance of 123.8. The average number of vehicles passing through the area per second was 474.6 with a variance of 11.6. The average speed of vehicles in this scenario was 5.22 m/s with a variance of 2.61. This scenario represents the morning rush hour, when traffic congestion is usually high as commuters head to work or school.

Scenario 2 occurred from 13:00 to 13:05, and a total of 862 vehicles were tracked. The average dwell time was 188.5 seconds with a variance of 125.1. The average number of vehicles passing through the area per second was 541.6 with a variance of 5.38. The average vehicle speed in this scenario was 5.59 m/s with a variance of 2.73. This scenario likely represents midday traffic, where there may be a mix of commuters, delivery vehicles, and other traffic on the road. Overall, these statistics provide valuable insights into the different traffic patterns and characteristics throughout the day, which is crucial for planning and predicting the resources required for the vehicle fog system.

Scenario 3, which occurred from 18:00 to 18:05, tracked a total of 928 vehicles in a $3 \times 3$ $km^2$ area in Qingyang District, Chengdu, China. The average vehicle dwell time in this scenario was 196.5 seconds with a variance of 122.5 seconds. The average number of vehicles passing through the area per second was 608.0 with a variance of 7.76. The average vehicle speed was 4.60 m/s, with a variance of 2.40 m/s. This scenario represents the evening rush hour traffic in this area, with a high volume of vehicles passing through at relatively lower speeds compared to the other scenarios.

Scenario 4 was tracked from 23:00 to 23:05 with a total of 359 vehicles in a $3 \times 3$ $km^2$ area. The average dwell time of vehicles in this scenario was 173.7 seconds with a variance of 124.1 seconds. The average number of vehicles passing through the area per second was 207.9 with a variance of 3.93. The average vehicle speed was 7.30 m/s, with a variance of 3.16 m/s. This scenario represents late-night traffic, with fewer vehicles on the road compared to other times of the day, but with a higher average speed. The lower variance in dwell time and number of vehicles indicates a more steady flow of traffic during this time period.

\subsection{Evaluation metrics}
To evaluate the performance, the following information is collected: the upload time and processing time of each task; the total number of tasks, denoted by $K_{total}$ and the number of tasks served, denoted by $K_{serviced}$. The total number of tasks executed locally, denoted by $K_{local}$. Accordingly, four metrics, namely average processing time (APT), average service time (AST), average service ratio (ASR), and cumulative reward (CR), are defined based on the following relationships.

Average processing time is a metric that calculates the average time taken to process each job. It is calculated by summing the processing time of each job and dividing it by the total number of jobs serviced.

\begin{equation}
APT=\frac{\sum_{i=1}^{n}Task\ processing\ time}{K_{serviced}}
\label{Eq11}
\end{equation}

Average service time is a metric that calculates the average time taken to complete each task. It includes both the processing time and the upload time of each task. It is calculated by summing the upload time and processing time of each task and dividing it by the total number of tasks serviced.
\begin{small}
\begin{equation}
AST=\frac{\sum_{i=1}^{n}\left( Task\ processing\ time+upload \ time \right)}{K_{serviced}}
\label{Eq12}
\end{equation}
\end{small}
The average service ratio is a measurement that calculates the ratio of the number of tasks serviced to the total number of tasks. This indicates the effectiveness of the system in performing tasks.

\begin{equation}
ASR=\frac{k_{serviced}}{k_{total}}
\label{Eq13}
\end{equation}

Cumulative reward is a measure that evaluates the overall performance of the system. It is obtained by combining rewards from different criteria such as minimizing resource waste, maximizing resource utilization, response time, and optimal quality of service.
\begin{small}
\begin{equation}
\begin{split}
CR = &\ \left( \min(Resource_{wastage}) + \max(Resource_{utilization}) \right. \\
&\ \left. + \min(Response\text{-}time) + Quality \ of \ Service \right)
\label{Eq14}
\end{split}
\end{equation}
\end{small} 

In order for different types of values and units to be summed together, they must first be normalized to a common scale. This can be achieved by assigning a weight to each criterion based on its importance, and then converting the values for each criterion to a standard score between 0 and 1. Once all the criteria have been normalized, they can be summed together to obtain an overall CR score.

Average achieved potential (AAP): It is defined as the sum of edge rewards (potential gained) over the number of edge nodes during the scheduling period, which is calculated by the Eq. \ref{Eq15}.
\begin{equation}
\frac{1}{E} \sum_{i=0}^{n}e \ \forall e \in E + \sum_{i=0}^{n}t \ \forall t \in T_{e}^{r^t}
\label{Eq15}
\end{equation}

In Eq. \ref{Eq15}, $E$ represents the set of all edges in a network. An edge is a connection or link between two nodes in a network. $e$ is a distinct edge of the set $E$. The Equation is repeated on each edge $e\in E$. $r$ represents the edge reward. Each edge has a reward associated with it, which can be a numerical value that represents the potential or value of that edge. $t$ is a specific time period at the edge node. The equation is repeated at each time period $t \in T_e$, where$T_e$ denotes the set of all time periods associated with edge $e$. $T$ represents the set of all time periods in which the computation is performed. Each edge has a set of time periods $T_e$ associated with it.

Edges are the edges or connections between nodes in a network. In other words, edges represent possible connections between different components of a network. If the reward of an edge is high, it may indicate that the connection between two nodes in the network is very valuable and that this connection is used effectively for the intended purpose. On the other hand, if the reward of an edge is low, it may indicate that the connection between two nodes in the network is of little value. In proximal policy optimization computation, the reward value of an edge influences the decisions based on which connections between nodes are considered.

Edge reward refers to the additional reward or incentive given to an edge in a network. Edge reward is a numerical value that represents the potential or additional value of that edge. Increasing the edge reward provides the advantage of attracting more traffic or load to that particular edge and encourages the use and prioritization of that edge for communications and data transmission. This can help optimize resource utilization and improve overall system performance by reducing congestion and minimizing resource waste. On the other hand, decreasing the edge reward leads to less traffic or load being directed to that edge. This can be beneficial if there are other edges with better capabilities or resources that should be prioritized. By decreasing the edge reward, the system can distribute the load more evenly or direct it to more optimal paths or edges.

\subsection{Simulation environment}
Table \ref{table4} depicts the simulation environment.
The proposed method is implemented using Python and Matlab 2020a. The model is run on an Ubuntu 20.04 server with a 16-core AMD Ryzen 9 5950X processor clocked at 4.3 GHz, two NVIDIA GeForce RTX 3090 GPUs, and 64 GB of memory. The proposed model is run on all nodes of the fog layer.
{The main training parameters are configured as follows: learning rate ($\alpha$)=0.1, discount factor ($\lambda$)=0.9, and $\epsilon$-greedy exploration rate decaying from 0.1 to 0.01. The model is trained over 100 episodes. Fog nodes are deployed in quantities ranging from 5 to 10, featuring diverse CPU and memory configurations. This heterogeneity mirrors practical vehicular fog computing scenarios. All fog nodes start with low initial utilization (20\% CPU, 15\% memory) to simulate a baseline empty state, with queues initialized to zero.  }

\begin{table}[h]
\centering
\small
\caption{System parameter specifications}
\resizebox{0.5\textwidth}{!}{ 
\begin{tabular}{|p{6cm}|l|}
\toprule
\textbf{Parameter} & \textbf{Value} \\
\midrule
Task size &   [5, 10] MB  \\ 
 { Task compute requirement} &{100-500 MIPS} \\
Computing cycles for processing 1-bit work data &  500 cycles/bit  \\
Deadline for completing tasks & [5,10]sec  \\
V2I communication bandwidth & 20MHZ  \\
Edge node computation capability & [3,10] GHz  \\ 
Maximum power consumption of V2I communications & $10^3$mW  \\
V2I communication range & 500m \\
Wired transfer rate & 50Mbs \\
bAdditive white Gaussian noise & -90dBm \\
Noise power & -114dBm \\
User equipment bandwidth &20 MHZ \\
Path loss power in  large scale & 3dB \\
 {Simulation area} & {$3\times3$ $km^2$ }\\
{Learning rate ($\alpha$) }&{0.1} \\
{ Discount factor ($\lambda$)}&  { 0.9} \\
{Episodes} & {100} \\
{$\epsilon-greedy$}&  {0.1 (decays to 0.01 over episodes) } \\
{Fog nodes  } & { 5–10 nodes  } \\
{Fog node cpu utilization}& {20\%} \\
{Fog node memory utilization}& {15\%} \\

\bottomrule
\end{tabular}
}
\label{table4}
\end{table}

Gigahertz refers to the frequency at which an edge node is capable of performing computations. This does not necessarily mean that the processor clock speed is 10 GHz. Computing power is not directly calculated from frequency, as the number of cores and other factors also play a role in determining the overall processing power of a system. Two processors with the same frequency but different numbers of cores can have different computing power. 1-bit work refers to processing a data unit of size 1 bit \cite{ref81}. {The processing power requirement of the tasks are between 100-500 MIPS}.The vehicle-to-infrastructure communication bandwidth, 20 MHz, refers to the data transmission frequency, not the actual size of the data being transmitted. To calculate the data transfer rate, Shannon's law can be used, which states that the maximum data transfer rate (in bits per second) in a channel with bandwidth $B$ and signal-to-noise ratio $S/N$ is given by:

\begin{equation}
R=B* log_2(1+\frac{S}{N})
\label{Eq16}
\end{equation}

The simulation is performed in a $3\times 3$ $km^2$ area, with 9 edge nodes consisting of 5G base stations and roadside processing units uniformly distributed in the area, which is in accordance with \cite{ref79,ref81,ref83,ref84}. According to \cite{ref79,ref81,ref82,ref84}, the {$3\times 3$ $km^2$} area provides an acceptable scale for simulating and evaluating vehicle resource allocation in a fog computing environment. This allows a sufficient number of edge nodes, in this case 9 edge nodes consisting of 5G base stations and roadside processing units, to be uniformly distributed to experiment with different resource allocation strategies. The $3 \times 3$ area provides a balance between realism and computational complexity. The computational capacities of the edge nodes, represented by their CPU clock frequencies, are varied and uniformly distributed in the range of 3–10 GHz according to \cite{ref83}. The communication range for vehicle-to-infrastructure communications is set at 500 meters \cite{ref83}.

The proposed work is compared to Round Robin (RR) \cite{ref80}, First Come First Served (FCFS)\cite{ref80}, Weighted Fair Queuing (WFQ) and work based in Lagrange \cite{ref81}.  The round robin algorithm is a basic resource allocation technique that ensures fair resource allocation among different vehicles in a vehicle fog computing environment. This algorithm allocates available resources to vehicles in a round robin manner, where each vehicle is given a fixed amount of resources sequentially. The allocation continues in a round robin manner until all vehicles receive their required resources. This algorithm works well when vehicles have similar resource requirements. The advantage of the RR algorithm is that each task is executed in turn and there is no need to wait for the previous task to complete. However, if the load is found to be heavy, the round robin takes a long time to complete all tasks. First come first serve algorithm is another traditional resource allocation method that can be applied to vehicle fog computing. In this algorithm, vehicles are serviced in the order in which they request resources. The first vehicle to request resources is allocated the available resources first, and subsequent vehicles allocate resources in the same order.  Weighted Fair Queuing scheduling algorithm used in network and resource management to ensure fair allocation of resources among fog nodes. A fog computing environment consists of multiple fog nodes that act as intermediate points between end devices and cloud servers. The aim of paper \cite{ref81} is to reduce the service time by allocating the available bandwidth. An application model is built considering the service methods and solved through a two-stage approach. For the first stage, all suboptimal solutions are presented based on the Lagrange algorithm. For the second stage, an optimal solution selection process is presented and analyzed.

\subsection{Effect of different traffic scenarios}

Figure \ref{fig2} compares the five algorithms under different traffic scenarios. In Figure \ref{fig2}(a), the ASR above shows that the proposed method successfully allocates resources to vehicles and ensures that they receive the resources they need. 
{As depicted in Figure \ref{fig2}(a), the average ASRs for the proposed approach are 78\%, while those for FCFS, \cite{ref83}, RR, and WFQ are 62\%, 68\%, 63\%, and 72\%, respectively. In Figure \ref{fig2}(a), the Average Service Ratio (ASR) of the proposed Q-Learning method consistently outperforms traditional baselines such as FCFS, RR, WFQ, and the Lagrange-based approach \cite{ref83} across all traffic scenarios, demonstrating superior task servicing even under varying densities and mobility patterns. Furthermore, the proposed method maintains stable ASR performance throughout the scenarios, highlighting its robustness in handling non-stationary traffic flows where baselines exhibit noticeable degradation due to their lack of predictive capabilities.}
This can be attributed to the efficient resource allocation strategy used in the proposed method that considers the needs and priorities of vehicles.

\begin{small}
\begin{figure}[htb!]
    \centering
    \begin{minipage}[t]{0.48\textwidth}   
        \begin{subfigure}{0.95\textwidth}
            \centering
            \includegraphics[width=\linewidth, height=0.25\textheight]{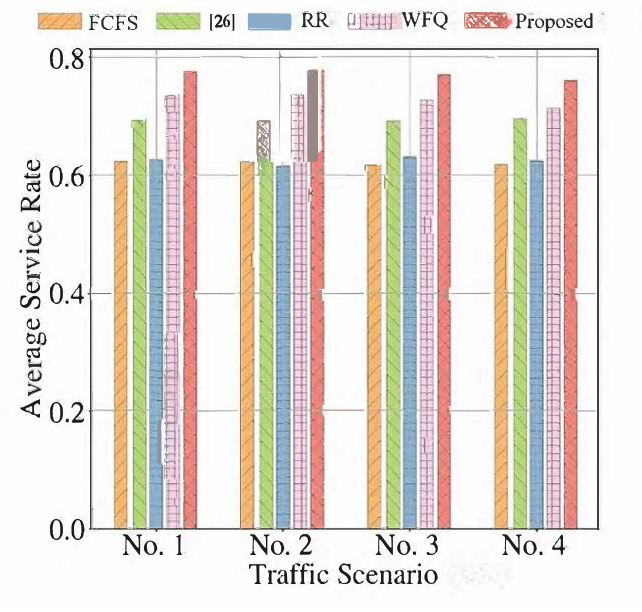}
            \caption*{(a) Average service rate.}
        \end{subfigure}
        
        \begin{subfigure}{0.95\textwidth}
            \centering
            \includegraphics[width=\linewidth, height=0.25\textheight]{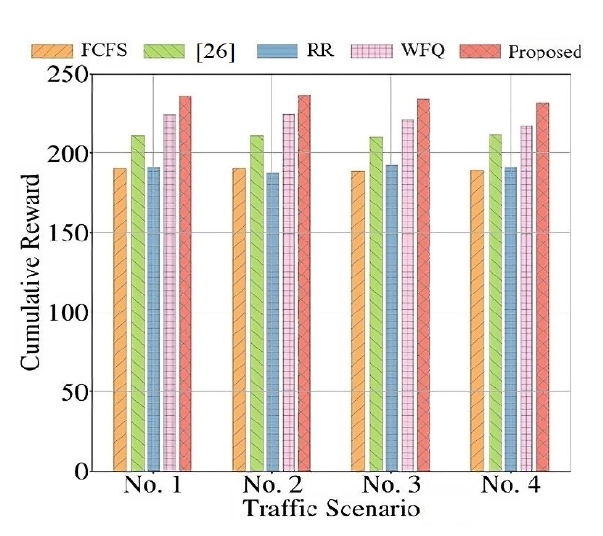}
            \caption*{(b) Cumulative reward.}
        \end{subfigure}
        
        \begin{subfigure}{0.95\textwidth}
            \centering
            \includegraphics[width=\linewidth, height=0.25\textheight]{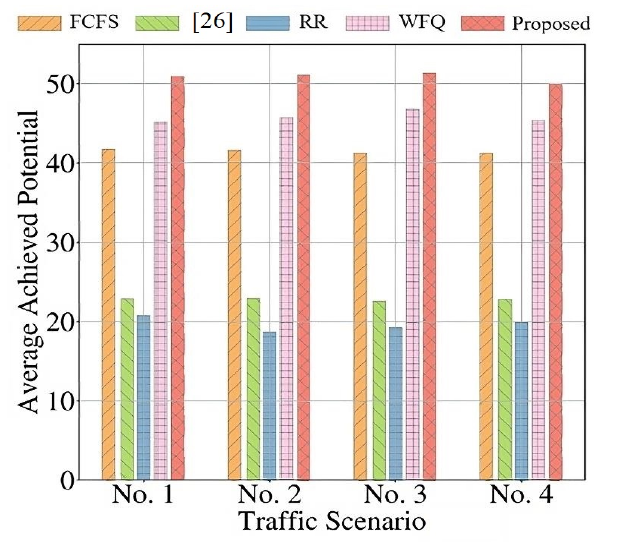}
            \caption*{(c) Average achieved potential.}
        \end{subfigure}

    \end{minipage}

    \caption{Performance comparison for different scenarios.}
    \label{fig2}
\end{figure}
\end{small}

\begin{small}
\begin{figure}[htb!]
    \centering
    \begin{minipage}[t]{0.48\textwidth} 
 \begin{subfigure}{0.95\textwidth}
            \centering
            \includegraphics[width=\linewidth, height=0.25\textheight]{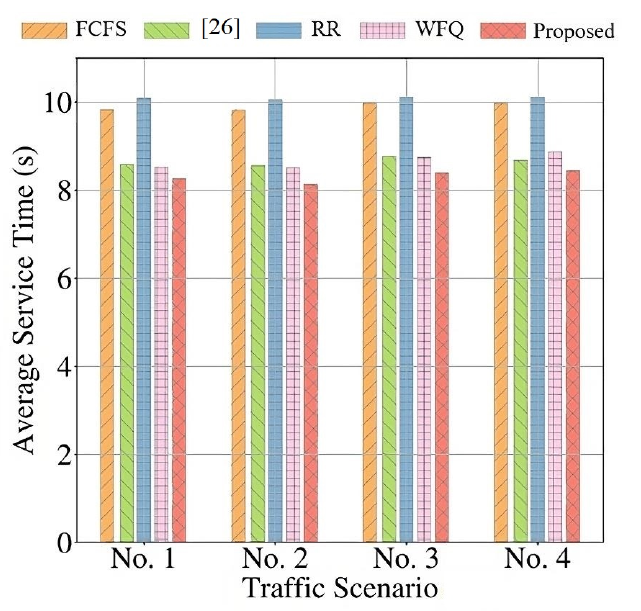}
            \caption*{(a) Average service time.}
        \end{subfigure}

           \begin{subfigure}{0.95\textwidth}
            \centering
            \includegraphics[width=\linewidth, height=0.25\textheight]{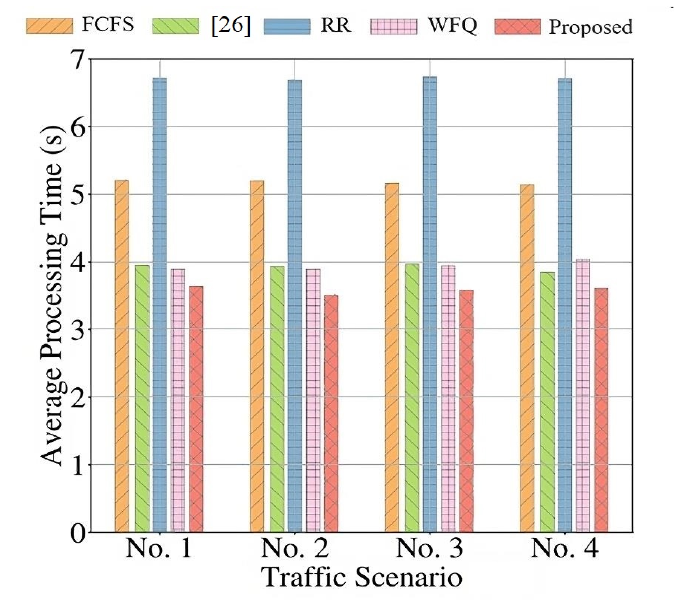}
            \caption*{(b) Average processing time.}
        \end{subfigure}

 \end{minipage}

    \caption{Performance comparison for different scenarios.}
    \label{fig3}
\end{figure}
\end{small}

 {Figure \ref{fig2} (b) depicts the cumulative reward for different approaches. The average CRs for the proposed approach are 235 for all scenarios, and the while those are 190, 212, 191, and 225 for FCFC, \cite{ref83}, RR, and WFQ, respectively. The proposed Q-Learning approach consistently demonstrates the highest reward levels in all scenarios, reflecting its superior ability to balance resource demands and adapt to varying traffic conditions, while traditional baselines like FCFS, RR, WFQ, and the Lagrange-based method yield progressively lower rewards, particularly in denser or more dynamic settings. This visual trend underscores the proposed method's robustness in maximizing long-term system efficiency, as it effectively learns from interactions to prioritize high-value actions, avoiding the inefficiencies seen in static or greedy strategies that struggle with real-time variability.} 

In Figure \ref{fig2} (c), the proposed method is expected to achieve the highest AAP under all scenarios.
 { Figure \ref{fig2} (c) depicts the average achieved poetical (AAP) for different approaches. The average AAPs for the proposed approach are 51 for all scenarios, while those are 42, 23, 20, and 46 for FCFC, \cite{ref83}, RR, and WFQ, respectively. The proposed Q-Learning approach leads with the highest value of AAP in every scenario, showing its consistent excellence in realizing system potential under diverse conditions, while baselines such as FCFS, RR, WFQ, and the Lagrange-based method \cite{ref83} show lower AAP, particularly struggling in denser or more variable traffic flows. This pattern highlights the proposed method's strength in adaptive learning, which maximizes resource efficiency and task fulfillment by responding fluidly to real-time demands, in contrast to the more rigid strategies of traditional approaches that falter amid changing dynamics.  }
 The higher AAP indicates that the proposed method allocates a higher proportion of available resources to vehicles and meets their resource needs more effectively. The weighting mechanism used in the method allows it to allocate resources more effectively based on the vehicle priorities.

{ Figures \ref{fig3} (a), and \ref{fig3} (b) depict the average service time (AST) and average processing time (APT) across four traffic scenarios for various resource allocation methods. The proposed Q-Learning approach shows the shortest bars in both metrics throughout the scenarios, indicating its consistent efficiency in delivering quick responses and computations under different conditions, while baselines like FCFS, RR, WFQ, and the Lagrange-based method display longer bars that grow taller in denser or more variable traffic, revealing their struggles with delays. This pattern highlights the proposed method's strength in adaptive learning, which smoothly handles real-time demands to minimize wait times, in contrast to the more rigid strategies of traditional approaches that often extend processing under pressure.  }

 The superior performance of the proposed method can be attributed to the use of QL reinforcement learning, which optimizes resource allocation based on vehicle requirements and priorities. This approach enables better resource utilization, faster task processing, and efficient coverage of vehicle resource needs. In contrast, other algorithms do not perform well due to their limitations. The FCFS algorithm lacks consideration of resource needs and priorities, which leads to potentially inefficient resource allocation. The rotating turn algorithm does not dynamically adapt to the different resource needs of vehicles and can lead to unequal resource utilization. The weighted fair queuing algorithm used in \cite{ref83} does not sufficiently consider vehicle priorities, which potentially leads to suboptimal resource allocation. Furthermore, \cite{ref83} uses the Lagrangian algorithm, which does not obtain optimal solutions. Overall, the proposed method with reinforcement learning allows it to consider resource requirements, priorities, and achieve efficient resource allocation, thus outperforming other algorithms. This shows that, as expected, the proposed method can achieve shared communication and computation between edge nodes and improve the overall service ratio by minimizing the average service time of tasks.

\subsection{The effect of task arrival probability}

The effect of task arrival probability refers to how the frequency of sending tasks to the fog system affects its performance. In the proposed method, the performance of five algorithms under different vehicle arrival probabilities is analyzed and compared. This analysis helps to understand how different levels of task arrival rates affect resource scheduling and system forecast accuracy. By studying the effect of task arrival probabilities, the optimal strategy for efficient resource allocation in the fog system can be determined.

\begin{small}
\begin{figure}[htb!]
    \centering
    \begin{minipage}[t]{0.48\textwidth}   
        \begin{subfigure}{0.95\textwidth}
            \centering
            \includegraphics[width=\linewidth, height=0.25\textheight]{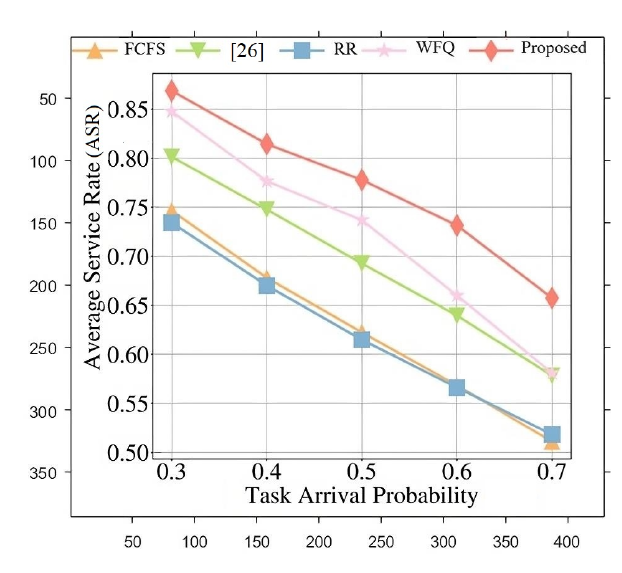}
            \caption*{(a) Average service rate.}
        \end{subfigure}
        
        \begin{subfigure}{0.95\textwidth}
            \centering
            \includegraphics[width=\linewidth, height=0.25\textheight]{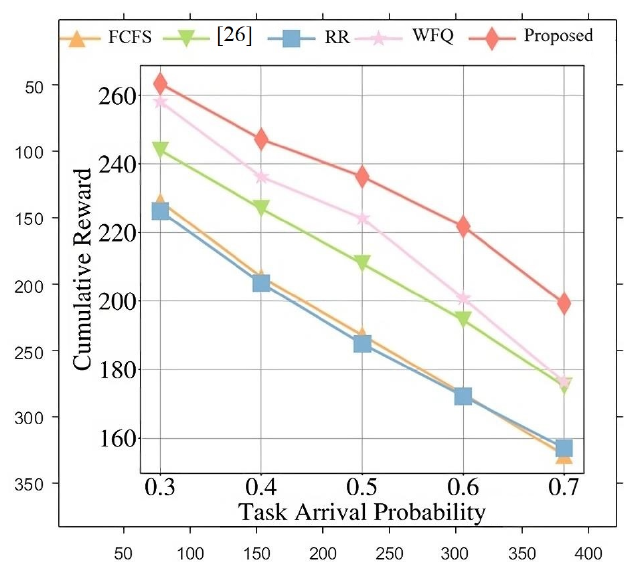}
            \caption*{(b) Cumulative reward.}
        \end{subfigure}
        
        \begin{subfigure}{0.95\textwidth}
            \centering
            \includegraphics[width=\linewidth, height=0.25\textheight]{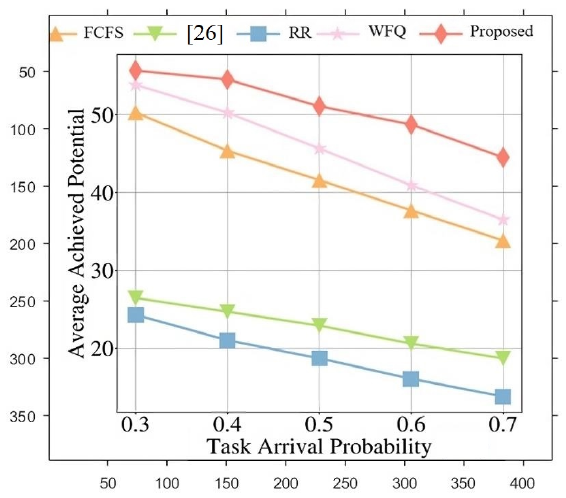}
            \caption*{(c) Average achieved potential.}
        \end{subfigure}

    \end{minipage}

    \caption{Performance comparison under different task entry probabilities.}
    \label{fig4}
\end{figure}
\end{small}

\begin{small}
\begin{figure}[htb!]
    \centering
    \begin{minipage}[t]{0.48\textwidth} 
 \begin{subfigure}{0.95\textwidth}
            \centering
            \includegraphics[width=\linewidth, height=0.25\textheight]{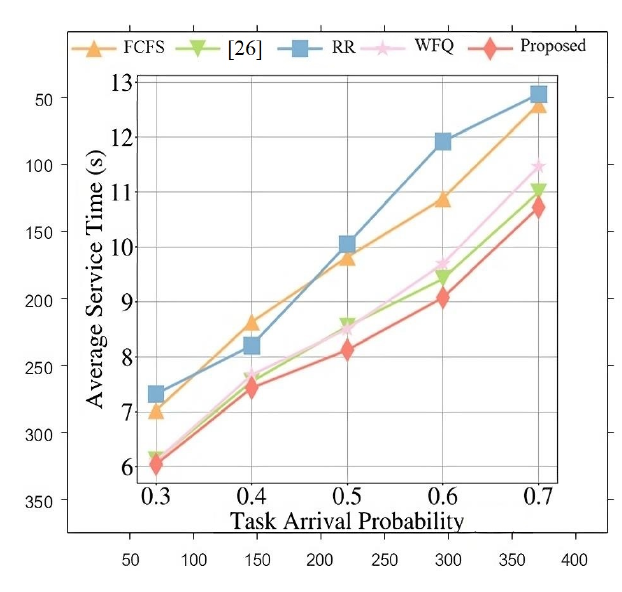}
            \caption*{(a) Average service time.}
        \end{subfigure}

           \begin{subfigure}{0.95\textwidth}
            \centering
            \includegraphics[width=\linewidth, height=0.25\textheight]{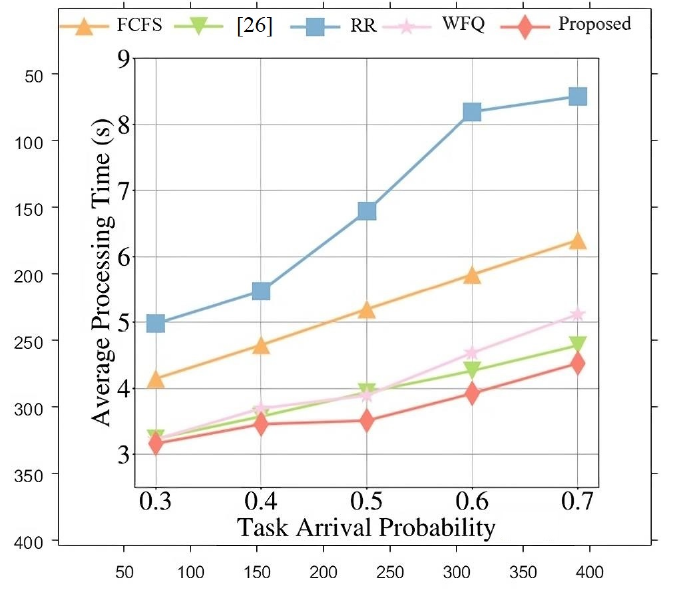}
            \caption*{(b) Average processing time.}
        \end{subfigure}

 \end{minipage}

    \caption{Performance comparison under different task entry probabilities.}
    \label{fig5}
\end{figure}
\end{small}

Figures \ref{fig4}, and \ref{fig5} compare the five algorithms under different vehicle arrival probabilities. In this set of experiments, we consider that the vehicle arrival probability increases from 3\% to 5\% in each time interval. It is expected that the performance of all five algorithms deteriorates as the vehicle arrival probability increases.

Figure \ref{fig4}(a) compares the ASR of the five algorithms and the proposed method achieves the highest ASR. As the probability of vehicle arrival increases, the service rate decreases because with more vehicles arriving, the system becomes busier and servers handle more tasks simultaneously. This can lead to congestion, delay, and increased waiting time for processing each task. As a result, the average service time increases, which leads to a decrease in the service rate. In addition, servers may become overwhelmed and cannot process tasks as efficiently, which further affects the service rate. Figures \ref{fig4}(b) and \ref{fig4}(c) also compare the CR and AAP for the five algorithms and show that the proposed method can have the highest CR and AAP, which indicates the advantages of the proposed method by adopting the third-season functions as the edge node reward.

{Figure \ref{fig4} (b) delves into cumulative reward (CR), revealing the proposed method's marked dominance with rewards peaking at approximately 50 at low average arrival probabilities $(AAP \approx 0.3)$, gradually tapering to around 20 at AAP = 0.7, yet maintaining a steeper upward curve compared to baselines like FCS (which plateaus earlier) and WFQ (showing sharper drops), underscoring the proposed algorithm's robustness in maximizing long-term value accumulation even under intensified workloads. Similarly, Figure \ref{fig4}(c) examines average achieved potential (AAP), where the proposed method secures the highest values—reaching nearly 35 at AAP = 0.3 and sustaining above 25 at higher loads, while competitors such as RR and \cite{ref83} falter with more pronounced declines, emphasizing the innovative integration of adaptive scheduling and resource allocation in the proposed framework, which not only mitigates overload but also enhances overall system potential realization by optimizing task prioritization and server utilization. }

Figures \ref{fig5} (a), and \ref{fig5} (b) compares the five algorithms (AST and APT). It can be seen that the performance gap between WFQ and the proposed method in this paper \cite{ref83} is small when the probability of the job arriving increases from 0.3 to 0.4. This is because the scheduling effect is not significant when there are sufficient resources.
 { In Figures \ref{fig5} (a) and \ref{fig5}(b), the comparison across the five algorithms reveals distinct trends in average service time and average processing time under varying task entry probabilities. The proposed method generally demonstrates superior performance, maintaining lower times compared to baselines like FCS, \cite{ref83}, RR, and WFQ, particularly as arrival rates intensify, where more advanced scheduling shines by efficiently allocating resources to minimize delays. Notably, the gap between WFQ and the proposed approach narrows at lower to moderate arrival probabilities, as abundant system resources diminish the impact of sophisticated scheduling, allowing simpler methods to perform comparably without significant congestion or resource contention.}

The proposed method outperforms other algorithms for several reasons:
\begin{enumerate}
\item Efficiency in resource allocation: The proposed method optimizes resource allocation based on a reward system considering the probability of task arrival. By assigning priority to vehicles with higher rewards, this method ensures efficient resource utilization and reduces waiting time and task delays.
\item Adaptation to changing conditions: As the probability of tasks arriving increases, the performance gap between the proposed method and the weighted fair queue remains small. This indicates that the proposed method effectively adapts to the changing workload by dynamically adjusting the resource allocation based on the reward system.
\item 3. Increased fairness: The weighting mechanism in the weighted fair queue may not consider the priority or importance of tasks.
In contrast, the proposed method assigns higher rewards to vehicles with higher resource requirements or critical tasks. This
ensures fair resource allocation and prevents low-priority tasks from affecting the performance of other vehicles.
\item Comprehensive evaluation criteria: The analysis considers ASR, CR, AAP, AST, and APT to evaluate the performance of the algorithms. The proposed method consistently achieves higher values for these criteria, demonstrating its effectiveness in resource allocation and task scheduling.
\end{enumerate}

 { While the proposed Q-learning-based method demonstrates superior performance in dynamic resource allocation for three-tier vehicular fog computing, it is limited by its dependency on high-quality input data for accurate predictions, the inherent complexity of the Q-learning algorithm which may lead to longer convergence times in highly volatile environments, and constraints imposed by limited hardware resources in vehicles, potentially exacerbating issues like data uncertainty and adaptation delays under extreme network congestion. Additionally, the approach assumes a relatively stable three-tier architecture, which may not fully account for real-world variabilities such as intermittent connectivity or heterogeneous vehicle capabilities, thereby restricting its scalability in ultra-large-scale deployments.}


\section{Conclusion}
\label{conclusion}

In this paper, a proposed method using QL reinforcement learning for managing and predicting the required resources of an vehicular fog system with a three-layer architecture was proposed in order to optimize resource allocation and improve the overall performance of the system. This study shows that the use of QL reinforcement learning is very effective in optimizing resource allocation, because the system can learn from past experiences, adapt to changing conditions, and make decisions based on real-time data.

One of the advantages of this method is its ability to handle dynamic and unpredictable situations, which, by using the QL reinforcement learning agent, can dynamically adjust resource allocation strategies based on the changing needs of the system. This adaptability increases the efficiency of the fog system, and by optimizing resource allocation, this scheme can potentially increase the overall reliability and availability of the automotive fog system, leading to improved user experience and satisfaction. The findings of this research can have practical implications for the development of more effective automotive fog computing architectures and resource management strategies that benefit both industry and academia.

Despite its high reliability and favorable performance in real-world environments, this method faces challenges, limitations, and weaknesses that require attention and improvement. These challenges include: Dependency to input data, hardware resource limitation in vehicles, complexity of Q-learning algorithm, data uncertainty, and adapting to changing conditions.
 { For future work, integrating hybrid reinforcement learning techniques, such as combining Q-learning with deep neural networks, could enhance convergence speed and robustness; conducting real-world pilot implementations in diverse urban scenarios would validate and refine the model; and exploring multi-objective optimization to balance energy efficiency alongside performance metrics would broaden its applicability in sustainable IoV ecosystems.} 

\bibliographystyle{unsrtnat}

\bibliography{vfog}
%
%
%
\end{document}